\documentstyle[12pt]{article}
\newlength{\extraspace}
\newlength{\extraspaces}
\newcommand{\be}{\begin{equation}
\addtolength{\abovedisplayskip}{\extraspaces}
\addtolength{\belowdisplayskip}{\extraspaces}
\addtolength{\abovedisplayshortskip}{\extraspace}
\addtolength{\belowdisplayshortskip}{\extraspace}}
\newcommand{\ee}{\end{equation}}

\newcommand{\ba}{\begin{eqnarray}
\addtolength{\abovedisplayskip}{\extraspaces}
\addtolength{\belowdisplayskip}{\extraspaces}
\addtolength{\abovedisplayshortskip}{\extraspace}
\addtolength{\belowdisplayshortskip}{\extraspace}}
\newcommand{\ea}{\end{eqnarray}}

\newcommand{\nonu}{\nonumber \\[.5mm]}
\newcommand{\A}{&\!\!\!}

\newcommand{\newsection}[1]{
\vspace{7mm} \pagebreak[3] \addtocounter{section}{1}
\setcounter{subsection}{0} \setcounter{footnote}{0}
\begin{center}
{\large {\bf \thesection. #1}}
\end{center}
\nopagebreak
\medskip
\nopagebreak \hspace{3mm}}

\begin{document}

\pagenumbering{arabic}

\begin{center}
{{\bf Charged Axially Symmetric Solution, Energy and Angular
Momentum in Tetrad Theory of Gravitation}}
\end{center}
\centerline{ Gamal G.L. Nashed}

\bigskip

\centerline{{\it Mathematics Department, Faculty of Science, Ain
Shams University, Cairo, Egypt }}

\bigskip
 \centerline{ e-mail:nashed@asunet.shams.edu.eg}

\hspace{2cm}
\\
\\
\\
\\
\\
\\
\\
\\

Charged axially symmetric solution  of the coupled gravitational
and electromagnetic fields in the tetrad theory of gravitation is
derived. The metric associated with this solution is an axially
symmetric metric which is characterized by three parameters
``$\,$the gravitational mass $M$, the charge parameter $Q$ and the
rotation parameter $a$". The parallel vector fields and the
electromagnetic vector potential are axially symmetric. We
calculate the total exterior energy. The energy-momentum complex
given by M\o ller in the framework of the Weitzenb$\ddot{o}$ck
geometry ``$\,${\it characterized by vanishing the curvature
tensor constructed from the connection of this geometry}" has been
used.  This energy-momentum complex is considered as a better
definition for calculation of energy and momentum than those of
general relativity theory. The energy contained in a sphere is
found to be consistent with pervious results which is shared by
its interior
 and exterior. Switching off the charge parameter, one finds that no
energy is shared by the exterior of the charged axially symmetric
solution. The components of the momentum density are also
calculated and  used to evaluate the angular momentum
distribution. We found no angular momentum contributes to the
exterior of the charged axially symmetric solution if zero charge
parameter is used.

\newpage
\begin{center}
\newsection{\bf Introduction}
\end{center}

The search for a consistence expression for the gravitating energy
and angular momentum of a self-gravitating distribution of matter
is undoubtedly a long-standing problem in general relativity. The
gravitational field does not possess the proper definition of an
energy momentum tensor and one usually defines some
energy-momentum and angular momentum as Bergmann \cite{BT} or
Landau-Lifschitz \cite{LL} which are pseudo-tensors. The
Einstein's general relativity can also be reformulated in the
context of  Weitzenb$\ddot{o}$ck (teleparallel) geometry. In this
geometrical setting the dynamical field quantities correspond to
orthonormal tetrad fields ${b^i}_\mu$  (i, $\mu$ are SO(3,1) and
space-time indices, respectively). The teleparallel geometry is a
suitable framework to address the notions of energy, momentum
 and angular momentum of any spacetime that admits a $3+1$
 foliation \cite{ MR}. Therefore, we consider the tetrad theory of
 gravitation.

M\o ller modified general relativity by constructing a new field
theory using the Weitzenb$\ddot{o}$ck geometry \cite{Mo8}. The aim
of this theory was to overcome the problem of energy-momentum
complex that appears in the Riemannian space \cite{M2}.  The field
equations in this new theory were derived from a Lagrangian which
is not invariant under local tetrad rotation. S$\acute{a}$ez
\cite{Se} generalized M\o ller theory into a scalar tetrad theory
of gravitation.  Meyer \cite{Me} showed that M\o ller theory is a
special case of Poincar$\acute{e}$ gauge theory \cite{HS,HNV}.

The tetrad theory  of gravitation  based on the geometry of
absolute parallelism \cite{PP}$\sim$\cite{AGP} can be considered
as the closest alternative to general relativity, and it has a
number of attractive features both from the geometrical and
physical viewpoints. Absolute parallelism is naturally formulated
by gauging spacetime translations and underlain by the
Weitzenb$\ddot{o}$ck geometry, which is characterized by the
metricity  condition and by the vanishing of the curvature tensor
(constructed from the connection of the Weitzenb$\ddot{o}$ck
geometry). Translations are closely related to the group of
general coordinate transformations which underlies general
relativity. Therefore, the energy-momentum tensor represents the
matter source in the field equation for the gravitational field
just like in general relativity.

The tetrad formulation of gravitation was considered by M\o ller
in connection with attempts to define the energy of gravitational
field \cite{Mo8,Mo2}. For a satisfactory description of the total
energy of an isolated system it is necessary that the
energy-density of the gravitational field is given in terms of
first- and/or second-order derivatives of the gravitational field
variables. It is well-known that there exists no covariant,
nontrivial expression constructed out of the metric tensor.
However, covariant expressions that contain a quadratic form of
first-order derivatives of the tetrad field are feasible. Thus it
is legitimate to conjecture that the difficulties regarding the
problem of defining the gravitational energy-momentum are related
to the geometrical description of the gravitational field rather
than are an intrinsic drawback of the theory \cite{Mj,MDTC}.

Kawai et. al. \cite{KT} assuming the tetrad fields to have the
form \[ {b^i}_\mu={\delta^i}_\mu+\displaystyle{a \over
2}l^kl_\mu-\displaystyle{Q^2 \over 2}m^km_\mu,\] and \[ \eta^{\mu
\nu}l_\mu l_\nu=0, \qquad \qquad \eta^{\mu \nu}m_\mu m_\nu=0,
\qquad \qquad  \eta^{\mu \nu}l_\mu m_\nu=0, \] were able to obtain
a charged Kerr metric solution in new general relativity. In
extended new general relativity also Kawai et. al. \cite{KST} have
examined axi-symmetric solutions of the gravitational and
electromagnetic field equations in vacuum from the point of view
of the equivalence principle.

Tolman,  Landau and Lifshitz (LL), Einstein
 and  M\o ller's definitions have been used by  Virbhadra
\cite{Vs,Vs1} to evaluate the energy distribution in Kerr-Newman
  spacetime up to the third order of the rotation parameter $a$.
   For the first three definitions, he found the
  same results, but for the fourth definition the energy was not the same.
  The same calculations have been done by Cooperstock et al.  beyond the
seventh order of $a$ \cite{CR}. Virbhadra  also \cite{Vk9}
investigated whether or not the energy-momentum complexes of
Einstein, (LL),  Papapetrou and Weinberg can give the same energy
distribution for the most general nonstatic spherically symmetric
metric. He found that the results obtained from these definitions
do not agree.  Nahmad-Achar and Schutz, however, generalizing a
method given by Persides \cite{NS} and found that the total energy
associated with Kerr spacetime is shared by its interior as well
as exterior of the Kerr black hole. Virbhadra \cite{VM}
calculated the angular momentum distribution of the Kerr-Newman
using the Landau and Lifshitz pseudotensor. Ahmed et al. \cite{AH}
calculated the energy distribution in a general spacetime which
include all the physically interesting spacetimes such as the
black hole spacetimes as well as the NUT spacetime. Rosen and
Virbhadra have shown that several energy-momentum complexes can
give results with high degree of consistency. For a non-flat
spacetimes they \cite{RV}  obtained  the same and acceptable
energy and momentum distributions .

According to the above discussion, it is clear that there is a
problem in using the definitions of the energy-momentum complexes
resulting from general relativity theory of gravitation. It is the
aim of the present work to derive an exact charged axially
symmetric solution in M\o ller's tetrad theory of gravitation for
the coupled gravitational and electromagnetic fields.  Using this
solution we  calculate the energy and angular momentum using the
superpotential method given by M\o ller \cite{Mo8} and Mikhail
et.al \cite{MWHL}. The formula of the calculated energy is found
to be shared by its interior as well as exterior and depend on the
three parameters (i.e., $M$, $Q$ and $a$). Switching off the
charge parameter $Q$ (i.e., $Q=0$) one finds that the energy will
give the correct formula of Schwarzschild or Kerr black hole. On
the other hand, if the rotation parameter $a=0$ then the formula
of the calculated energy will coincide with that of
 Reissner-Nordstr$\ddot{o}$m black hole. Since the axially
 symmetric solution describes the exterior field of a rotating charged object
therefore, there is an angular momentum distribution due to the
presence of the electromagnetic field which we have evaluated.

In \S 2 we derive the field equations of the coupled gravitational
and electromagnetic fields. In \S 3 we first apply the tetrad
field with sixteen unknown  functions of $\rho$ and $\phi$ to the
derived field equations. Solving the resulting partial
differential equations, we obtain an exact analytic solution. In
\S 4 we calculate the energy and angular momentum distribution of
this solution  keeping till the fourth order of the rotation
parameter $a$. The final section is devoted to discussion and
conclusion.
\newpage
\newsection{The tetrad theory of gravitation and electromagnetism}

In the Weitzenb{\rm $\ddot{o}$}ck spacetime the fundamental field
variables describing gravity are a quadruplet of parallel vector
fields \cite{HS} ${b_i}^\mu$\footnote{Latin indices
$(i,j,k,\cdots)$ designate the vector number, which runs from
$(0)$ to
 $(3)$, while Greek indices $(\mu,\nu,\rho, \cdots)$ designate the world-vector components
running from 0 to 3. The spatial part of Latin indices is denoted
by $(a,b,c,\cdots)$, while that of Greek indices by $(\alpha,
\beta,\gamma,\cdots)$.}, which we call the tetrad field in this
paper, characterized by \be D_{\nu} {b_i}^\mu=\partial_{\nu}
{b_i}^\mu+{\Gamma^\mu}_{\lambda \nu} {b_i}^\lambda=0\; , \ee where
${\Gamma^\mu}_{\lambda \nu}$ define the nonsymmetric affine
connection coefficients. The metric tensor $g_{\mu \nu}$
 is given by
 \[g_{\mu \nu}= \eta_{i j} {b^i}_\mu {b^j}_\nu\]
  with the
Minkowski metric $\eta_{i j}=\textrm {diag}(+1\; ,-1\; ,-1\;
,-1)$\footnote{ Latin indices are rasing and lowering with the aid
of $\eta_{i j}$ and $\eta^{ i j}$.}. Equation (1) leads to the
metricity  condition and the identically vanishing of curvature
tensor.

 The gravitational Lagrangian $L_G$ is an invariant constructed
 from $g_{\mu \nu}$ and the contorsion tensor
  $\gamma_{\mu \nu \rho}$ given by \be \gamma_{\mu \nu \rho} =
{b^i}_{\mu}b_{i \nu; \ \rho}  \;, \ee where the semicolon denotes
covariant differentiation with respect to Christoffel symbols. The
most general gravitational Lagrangian density invariant under
parity operation is given by the form \cite{HS,HN, MWHL}
 \be
{\cal
 L}_G  =  \sqrt{-g} L_G = \sqrt{-g} \left( \alpha_1 \Phi^\mu \Phi_\mu
 + \alpha_2 \gamma^{\mu \nu
\rho} \gamma_{\mu \nu \rho}+ \alpha_3 \gamma^{\mu \nu \rho}
\gamma_{\rho \nu \mu} \right) \ee
 with \[g = {\rm det}(g_{\mu
\nu})\] and $\Phi_\mu$ being the basic vector field defined by
 \be \Phi_\mu = {\gamma^\rho}_{\mu \rho}.\ee  Here $\alpha_1\; ,
\alpha_2\; ,$ and $\alpha_3$ are constants determined
 such that the theory coincides with general relativity in the weak
 fields \cite{Mo8,HN}:
\be
 \alpha_1=-{1 \over \kappa}\; , \qquad \alpha_2={\lambda \over
\kappa}\; , \qquad \alpha_3={1 \over \kappa}(1-\lambda)\; , \ee
 where
$\kappa$ is the Einstein constant and  $\lambda$ is a free
dimensionless parameter. The vanishing of this dimensionless
parameter will creates a teleparallel gravity equivalent to
general relativity \cite{PVZ}\footnote{ Throughout this paper we
use the relativistic units$\;$ , $c=G=1$ and
 $\kappa=8\pi$.}.

The electromagnetic Lagrangian  density ${\it L_{e.m.}}$ is
\cite{KT}
 \be {\it L_{e.m.}}=-\displaystyle{1 \over 4} g^{\mu \rho}
g^{\nu \sigma} F_{\mu \nu} F_{\rho \sigma}\; , \ee with $F_{\mu
\nu}$ being given by\footnote{Heaviside-Lorentz rationalized units
will be used.} $F_{\mu \nu}=
\partial_\mu A_\nu-\partial_\nu A_\mu$, where $A_\mu$ is the vector potential.

The gravitational and electromagnetic field equations for the
system described by ${\it L_G}+{\it L_{e.m.}}$ are the following:

 \be G_{\mu \nu} +H_{\mu \nu} =
-{\kappa} T_{\mu \nu}\; , \ee \be K_{\mu \nu}=0\; , \ee \be
\partial_\nu \left( \sqrt{-g} F^{\mu \nu} \right)=0 \ee
with $G_{\mu \nu}$ being the Einstein tensor of general
relativity.  Here
 $H_{\mu \nu}$ and $K_{\mu \nu}$ are defined by \be H_{\mu \nu}
= \lambda \left[ \gamma_{\rho \sigma \mu} {\gamma^{\rho
\sigma}}_\nu+\gamma_{\rho \sigma \mu} {\gamma_\nu}^{\rho
\sigma}+\gamma_{\rho \sigma \nu} {\gamma_\mu}^{\rho \sigma}+g_{\mu
\nu} \left( \gamma_{\rho \sigma \lambda} \gamma^{\lambda \sigma
\rho}-{1 \over 2} \gamma_{\rho \sigma \lambda} \gamma^{\rho \sigma
\lambda} \right) \right]\; ,
 \ee
and \be K_{\mu \nu} = \lambda \left[ \Phi_{\mu\; ,\nu}-\Phi_{\nu\;
,\mu} -\Phi_\rho \left({\gamma^\rho}_{\mu \nu}-{\gamma^\rho}_{\nu
\mu} \right)+ {{\gamma_{\mu \nu}}^{\rho}}_{;\rho} \right]\; , \ee
and they are symmetric and antisymmetric tensors, respectively.
The energy-momentum tensor $T^{\mu \nu}$ is given by \be T^{\mu
\nu}=-g_{\rho \sigma}F^{\mu \rho}F^{\nu \sigma}+\displaystyle{1
\over 4} g^{\mu \nu} F^{\rho \sigma} F_{\rho \sigma}. \ee

\newsection{ Exact Analytic Solution}

Let us begin with the tetrad field  which
 can be written in the spherical polar coordinates as

 \be \left({b^i}_{  \mu} \right)
= \left( \matrix{ A_1(\rho\; ,\phi) &   A_2(\rho\; ,\phi) &
A_3(\rho\; ,\phi) &  A_4(\rho\; ,\phi)
 \vspace{3mm} \cr  B_1(\rho\; ,\phi) \sin\theta
\cos\phi  & B_2(\rho\; ,\phi) \sin\theta \cos\phi & B_3(\rho\;
,\phi) \cos\theta \cos\phi
 & B_4(\rho\; ,\phi) \sin\phi  \sin\theta \vspace{3mm} \cr
  C_1(\rho\; ,\phi) \sin\theta \sin\phi  & C_2(\rho\; ,\phi)  \sin\theta \sin\phi &C_3(\rho\; ,\phi)
   \cos\theta
\sin\phi & C_4(\rho\; ,\phi)  \cos\phi \sin\theta \vspace{3mm} \cr
 D_1(\rho\; ,\phi) \cos\theta&   D_2(\rho\; ,\phi) \cos\theta & D_3(\rho\; ,\phi)\sin\theta  &
   D_4(\rho\; ,\phi) \cos\theta  \cr }
\right)\; , \ee where $A_i(\rho\; ,\phi)\; ,  \ B_i(\rho\;
,\phi)\; , \ C_i(\rho\; ,\phi) \ \ and \ \  D_i(\rho\; ,\phi)\; ,
\
 i=1 \cdots 4 $ are unknown functions of $\rho$  and  $\phi.$
Applying (13) to the field equations (7)$\sim$(9) one can obtains
a set of nonlinear partial differential equations. Due to the
lengthy of writing  these partial differential equations we will
just write the solution that satisfy  these differential
equations.
\vspace{.3cm}\\
\underline{The Exact Solution}\\

If the arbitrary functions take the following values \ba A_1 \A
=\A 1-\displaystyle{2M \rho-Q^2 \over 2 \Omega }\; , \quad
A_2=\displaystyle{2M \rho-Q^2   \over 2 \Upsilon}\; ,
 \quad A_3=0\; , \quad A_4=-\displaystyle{(2M \rho-Q^2) a \sin^2 \theta \over 2\Omega}\; , \nonu
 B_1 \A =\A \displaystyle{(2 M \rho-Q^2) \over 2 \Omega }\; ,
\quad B_2=\displaystyle {1 \over 2 \Upsilon \cos \Phi} \left(2
\rho \alpha- (2 M \rho-Q^2) \cos \Phi \right)\; ,\nonu
  B_3 \A =\A \displaystyle{\alpha \over \cos \Phi} \; ,
 \quad B_4=\displaystyle {-2 \beta+\displaystyle { (2 M \rho-Q^2) a \sin^2 \theta \cos\Phi
  \over \Omega} \over 2 \sin \Phi}\; , \nonu
C_1 \A =\A \displaystyle{ (2 M \rho-Q^2)  \over  2\Omega }\; ,
\quad C_2=\displaystyle {1 \over 2\Upsilon \sin \Phi}\left(2\rho
\beta-(2 M \rho-Q^2) \sin \Phi \right)\; ,\nonu
  C_3\A =\A \displaystyle{\beta \over \sin \Phi}\; , \quad
 C_4=\displaystyle {2 \alpha+\displaystyle { (2 M \rho-Q^2)a \sin^2 \theta \sin\Phi
  \over \Omega} \over 2 \cos \Phi}\; , \nonu
  D_1 \A =\A \displaystyle{(2 M \rho-Q^2)  \over  2 \Omega }\; ,
\quad D_2=1+\displaystyle{(2 M \rho-Q^2) \over   2 \Upsilon }\;
,\nonu
 D_3 \A=\A -\rho\; , \quad D_4=\displaystyle {(2 M \rho-Q^2) a \sin^2 \theta
  \over 2 \Omega}\; ,
 \ea
 where $\Omega, \ \  \Upsilon, \ \ \alpha, \ \ \beta \ \  and \ \ \Phi$ are defined by
  \ba \Omega \A =\A \rho^2+a^2 \cos^2\theta\; , \qquad \qquad  \Upsilon=\rho^2+a^2-2M\rho+Q^2
  \; , \nonu
\alpha \A=\A \rho \cos \Phi+a \sin\Phi \qquad \qquad \beta=\rho
\sin \Phi-a \cos \Phi\; , \qquad \qquad  \Phi=\phi-a \displaystyle
\int{d\rho \over \Upsilon}\; ,
  \ea
  $M$, $Q$ and $a$ are the gravitational mass, the charge
  parameter and the angular momentum of the rotating source \cite{Tn,KT}.

   The parallel vector fields (13) using the solution (14) is axially symmetric in
the sense that it is form invariant under the transformation \ba
\A \A \bar{\phi}\rightarrow \phi+\delta \phi\; , \qquad
\bar{b^{(0)}}_\mu \rightarrow {b^{(0)}}_\mu\; , \qquad
\bar{b^{(1)}}_\mu \rightarrow {b^{(1)}}_\mu \cos\delta
\phi-{b^{(2)}}_\mu \sin\delta \phi\; ,\nonu
\A \A \bar{b^{(2)}}_\mu \rightarrow {b^{(1)}}_\mu \sin \delta
\phi+{b^{(2)}}_\mu \cos \delta \phi \; , \qquad \bar{b^{(3)}}_\mu
\rightarrow {b^{(3)}}_\mu . \ea

 The form of  the antisymmetric electromagnetic
 tensor field  $F_{\mu \nu}$ and the energy-momentum tensor are given by
  \ba
  {\bf F} = \A \A
  \displaystyle{Q  \over 2\sqrt{\pi} \Omega^2} \Biggl\{ \left((a^2\cos^2 \theta-\rho^2) d\rho
  +2\rho a^2 \cos \theta \sin\theta d \theta \right)\wedge dt \nonu
  \A \A+ \left(2 \rho a \cos \theta \sin\theta(\rho^2+a^2) d\theta
  - a  \sin^2\theta(\rho^2-a^2\cos^2\theta)
    d\rho\right) \wedge d\phi \Biggr\}\; ,
\ea
  \ba
{T_1}^1 \A=\A -{T_2}^2=-\displaystyle{Q^2 \over 8 \pi \Omega^2}\;
,\nonu
  {T_3}^3 \A=\A -{T_0}^0 =\displaystyle{Q^2(\rho^2+a^2+a^2\sin^2 \theta) \over
8 \pi \Omega^3} \; , \nonu
{T_0}^3 \A=\A \displaystyle{-{T_3}^0 \over \sin^2
\theta(\rho^2+a^2)}  =\displaystyle{Q^2 a\over 4 \pi \Omega^3} .
  \ea
This solution satisfies the field equations (7)$\sim$(9) and the
associated metric has the following form \ba ds^2=\A \A
\left(\displaystyle{ \Upsilon-a^2\sin^2 \theta \over
\Omega}\right) dt^2 -\displaystyle{\Omega \over \Upsilon} d\rho^2
-\Omega d\theta^2- \displaystyle{\left \{ (\rho^2+a^2)^2-\Upsilon
a^2 \sin^2 \theta \right \} \over \Omega } \sin^2 \theta
d\phi^2\nonu
\A \A -2 \displaystyle{(2 M \rho-Q^2) a \sin^2 \theta \over
\Omega} dt d\phi
 \; ,\ea which is an axially symmetric metric.

The previously obtained solutions (Schwarzschild, Reissner
Nordstr$\ddot{o}$m and Kerr spacetimes) can be generated as
special solutions of solution (14) by an appropriate choice of the
arbitrary functions. Therefore, we are interested in this solution
 to calculate its associated exterior energy and its angular
momentum using the definition of the energy-momentum complex given
by M\o ller \cite{Mo8} within {\it the  framework of Weitzenb{\rm
$\ddot{o}$}ck spacetime}. For this purpose we will transform the
tetrad (13) using  solution (14) into the Cartesian form.

Performing the following coordinate transformation \cite{KT} \ba
\A \A t \rightarrow t-M\ln \Upsilon-2M^2 \left(1-\displaystyle{Q^2
\over 4M^2}\right) \int{\displaystyle{d\rho \over
\Upsilon_{KN}}}\; , \nonu
\A \A x \rightarrow (\rho \cos \Phi+a \sin \Phi) \sin \theta\; ,
\nonu
\A \A y \rightarrow (\rho \sin \Phi-a \cos \Phi) \sin \theta\; ,
\nonu
\A \A z \rightarrow \rho \cos \theta\; , \ea where \be
r=\sqrt{x^2+y^2+z^2}=\displaystyle{\sqrt{\rho^4+a^2\rho^2-a^2z^2}
\over \rho}\; , \ee to the tetrad (13) one can obtains\footnote{We
will denote the symmetric part by ( \ ), for example$\;$ ,
$A_{(\mu \nu)}=(1/2)( A_{\mu \nu}+A_{\nu \mu})$ and the
antisymmetric part by the square bracket [\ ], $A_{[\mu
\nu]}=(1/2)( A_{\mu \nu}-A_{\nu \mu})$ .} \ba \A  \A {b^{(0)}}_0 =
1-{(2M \rho-Q^2)\rho^2 \over 2\rho_1}\; , \nonu
 \A \A {b^{(0)}}_\alpha = \left\{   -(n_\alpha-\displaystyle {a \over \rho}
 \epsilon_{\alpha  j 3 }\;  n^j)
 -\displaystyle {a^2 \over \rho^2}  \displaystyle {z \over \rho} {\delta_\alpha}^3
  \right\}{(2M \rho-Q^2)\rho^4  \over 2\rho_1(\rho^2+a^2)}=-{b^{(l)}}_0 \; ,    \nonu
\A  \A  {b^{(l)}}_\beta  ={\delta^l}_\beta+ \Biggl \{x^l x_\beta
-2\displaystyle {a \over \rho} \epsilon_{k 3 (\beta} x^{l)} x^k +
\displaystyle {a^2 \over \rho^2} \left[ {\epsilon_{k}}^{l 3}
\epsilon_{m \beta 3} x^k x^m + 2 z\{\rho x^{(l}-\displaystyle{a
\over \rho} {\epsilon_{k 3}}^{(l} \; x^k \}{\delta_{\beta)}}^3
\right] \nonu
\A \A +\displaystyle{a^4 \over \rho^4}z^2 \delta_{\beta 3}
\delta^{3 l} \Biggr \} {(2M \rho-Q^2) \rho^4 \over 2\rho_1
(\rho^2+a^2)^2} \; , \ea where \be \rho_1=\rho^4+a^2z^2\; , \ee
and  the $\epsilon_{\alpha \beta \gamma}$ are the three
dimensional totally antisymmetric tensor with $\epsilon_{1 2
3}=1$.

\newsection{Energy and angular momentum associated with the axially symmetric solution}

 The superpotential is given by \cite{Mo8, MWHL}
  \be {{\cal U}_\mu}^{\nu \lambda} ={(-g)^{1/2} \over
2 \kappa} {P_{\chi \rho \sigma}}^{\tau \nu \lambda}
\left[\Phi^\rho g^{\sigma \chi} g_{\mu \tau}
 -\lambda g_{\tau \mu} \gamma^{\chi \rho \sigma}
-(1-2 \lambda) g_{\tau \mu} \gamma^{\sigma \rho \chi}\right]\; ,
\ee where ${P_{\chi \rho \sigma}}^{\tau \nu \lambda}$ is \be
{P_{\chi \rho \sigma}}^{\tau \nu \lambda} \stackrel{\rm def.}{=}
{{\delta}_\chi}^\tau {g_{\rho \sigma}}^{\nu \lambda}+
{{\delta}_\rho}^\tau {g_{\sigma \chi}}^{\nu \lambda}-
{{\delta}_\sigma}^\tau {g_{\chi \rho}}^{\nu \lambda} \ee with
${g_{\rho \sigma}}^{\nu \lambda}$ being a tensor defined by \be
{g_{\rho \sigma}}^{\nu \lambda} \stackrel{\rm def.}{=}
{\delta_\rho}^\nu {\delta_\sigma}^\lambda- {\delta_\sigma}^\nu
{\delta_\rho}^\lambda. \ee The energy-momentum density is defined
by \cite{Mo8} \be {\tau_\mu}^\nu={{{\cal U}_\mu}^{\nu
\lambda}}_{\; , \ \lambda},\ee where comma denotes ordinary
differentiation.  The energy $E$ contained in a sphere with radius
$R$ is expressed by the volume integral
   \cite{M58} \be
P_\mu(R)=\int_{r=R} \int \int {{{\cal U}_\mu}^{0 \alpha}}_{, \
\alpha} d^3 x=\int_{r=R} \int \int {\tau_\mu}^0 d^3x, \ee with
$P_0(R)=E(R)$ which is the energy and $P_\alpha(R)$ is the spatial
momentum.  For convenience of the calculations, we work for small
values of the rotation parameter $a$, so we will neglect the terms
beyond its fourth order. With this approximation, we have the
covariant components of the parallel vector fields (22) as
\\
\\
\ba {b^{(0)}}_0 \A=\A 1-\displaystyle{1 \over
r}\left(M-\displaystyle{Q^2 \over 2r}\right)-\displaystyle{a^2
\over 2r^3}\left[\left(M-\displaystyle{Q^2 \over
r}\right)-\left(3M-\displaystyle{2Q^2 \over
r}\right)\displaystyle{z^2 \over r^2}\right]\nonu
\A \A -\displaystyle{a^4 \over
8r^5}\left[\left(3M-\displaystyle{4Q^2 \over r}
\right)-\left(30M-\displaystyle{24Q^2 \over r}
\right)\displaystyle{z^2 \over r^2}+\left(35M-\displaystyle{24Q^2
\over r} \right)\displaystyle{z^4 \over r^4} \right],\nonu
 {b^{(0)}}_\alpha \A=\A -\left(M-\displaystyle{Q^2 \over 2r}\right)\displaystyle{n_\alpha \over
r}+\displaystyle{a \over r^2}\left(M-\displaystyle{Q^2 \over
2r}\right)\epsilon_{\alpha \beta 3} n^\beta + \displaystyle{a^2
\over r^3}\Biggl[\left\{2\left(M-\displaystyle{Q^2 \over
2r}\right)\displaystyle{z^2 \over r^2}+\displaystyle{Q^2 \over
4r^2}(x^2+y^2)\right\}n_\alpha\nonu
\A \A -\left(M-\displaystyle{Q^2 \over 2r} \right)\displaystyle{z
\over r}{\delta_\alpha}^3\Biggr]+\displaystyle{a^3 \over
2r^4}\left[\left(M-\displaystyle{Q^2 \over
r}\right)-\left(5M-\displaystyle{3Q^2 \over
r}\right)\displaystyle{z^2 \over r^2}\right]\epsilon_{\alpha \beta
3} n^\beta+\displaystyle{a^4 \over
8r^5}\Biggl[3\Biggl\{\displaystyle{Q^2 \over
2r}+\left(8M-\displaystyle{7Q^2 \over r}\right) \displaystyle{z^2
\over r^2} \nonu
\A \A -\left(16M-\displaystyle{21Q^2 \over
2r}\right)\displaystyle{z^4 \over r^4}\Biggr\}n_\alpha
-2\left\{2\left(2M-\displaystyle{3Q^2 \over
2r}\right)-\left(12M-\displaystyle{7Q^2 \over
r}\right)\displaystyle{z^2 \over r^2}\right\}\displaystyle{z \over
r} {\delta_\alpha}^3 \Biggr]=-\delta_{l \alpha}
{b^{(\alpha)}}_0,\nonu
 {b^{(l)}}_\alpha
\A=\A{\delta^l}_\alpha+\left(\displaystyle{n^l n_\alpha \over
r}-\displaystyle{2a \over r^2}{\epsilon^{(l}}_{\beta
3}n_{\alpha)}n^\beta \right)\left(M-\displaystyle{Q^2 \over
2r}\right)-\displaystyle{a^2 \over r^3}\Biggl[\displaystyle{1
\over 2}\left\{M+\left(5M-\displaystyle{3Q^2 \over
r}\right)\displaystyle{z^2 \over r^2}\right\}n^l n_\alpha \nonu
\A \A -\left(M-\displaystyle{Q^2 \over
2r}\right){\epsilon^l}_{\beta 3} {\epsilon_{\alpha \gamma}}^
3n^\beta n^\gamma-\left(M-\displaystyle{Q^2 \over
2r}\right)\displaystyle{2z \over r}n^{(l} {\delta_{\alpha)}}^3
\Biggr] +\displaystyle{a^3 \over
r^4}\Biggl[\Biggl\{\displaystyle{6z^2 \over
r^2}\left(M-\displaystyle{Q^2 \over 2r}\right)\nonu
\A \A+\displaystyle{Q^2 \over
2r^3}\left(x^2+y^2\right)\Biggr\}n^\beta{\epsilon^{(l}}_{\beta
3}n_{\alpha)}-\left(M-\displaystyle{Q^2 \over
2r}\right)\displaystyle{2z \over r}n^\beta {\epsilon^{(l}}_{\beta
3}{\delta_{\alpha)}}^3\Biggr]-\displaystyle{a^4 \over 8r^5}
\Biggl[\Biggl\{\left(5M-\displaystyle{4Q^2 \over r}
\right)-\Biggl(14M\displaystyle{z^2 \over r^2}\nonu
\A \A +\left(63M-\displaystyle{40Q^2 \over r} \right)
\displaystyle{z^4 \over r^4} \Biggr) \Biggr\}n^l
n_\alpha-8\left\{\left(M-\displaystyle{Q^2 \over
r}\right)-\left(7M-\displaystyle{4Q^2 \over r} \right)
\displaystyle{z^2 \over r^2}\right\} \displaystyle{2z \over
r}n^{(l}{\delta_{\alpha)}}^3-4\Biggl\{\left(M-\displaystyle{Q^2
\over r}\right)\nonu
\A \A -\left(8M-\displaystyle{5Q^2 \over r} \right)
\displaystyle{z^2 \over r^2}+\left(7M-\displaystyle{4Q^2 \over r}
\right) \displaystyle{z^4 \over r^4}\Biggr\}{\delta^l}_\alpha
 +4\left\{\left(M-\displaystyle{Q^2 \over
r}\right)-\left(9M-\displaystyle{5Q^2 \over r} \right)
\displaystyle{z^2 \over r^2}\right\} \delta^{3l} \delta_{3 \alpha}
\Biggr],
 \ea
and the contravariant components of the parallel vector fields
(22) have the form

 \ba {b_{(0)}}^0
\A=\A 1+\displaystyle{1 \over r}\left(M-\displaystyle{Q^2 \over
2r}\right)+\displaystyle{a^2 \over
2r^3}\left[\left(M-\displaystyle{Q^2 \over
r}\right)-\left(3M-\displaystyle{2Q^2 \over
r}\right)\displaystyle{z^2 \over r^2}\right]\nonu
\A \A +\displaystyle{a^4 \over
8r^5}\left[\left(3M-\displaystyle{4Q^2 \over r}
\right)-\left(30M-\displaystyle{24Q^2 \over r}
\right)\displaystyle{z^2 \over r^2}+\left(35M-\displaystyle{24Q^2
\over r} \right)\displaystyle{z^4 \over r^4} \right],\nonu
{b_{(0)}}^\alpha \A= \A {b^{(0)}}_\alpha= -\delta^{l \alpha}
{b_{(\alpha)}}^0 , \nonu
 {b_{(l)}}^\alpha
\A = \A{\delta_l}^\alpha-\left(\displaystyle{n_l n^\alpha \over
r}-\displaystyle{2a \over r^2}\epsilon_{\beta 3
(l}n^{\alpha)}n^\beta \right)\left(M-\displaystyle{Q^2 \over
2r}\right)+\displaystyle{a^2 \over r^3}\Biggl[\displaystyle{1
\over 2}\left\{M+\left(5M-\displaystyle{3Q^2 \over
r}\right)\displaystyle{z^2 \over r^2}\right\}n_l n^\alpha \nonu
\A \A -\left(M-\displaystyle{Q^2 \over 2r}\right)\epsilon_{l \beta
3} {\epsilon_{\gamma}}^{3 \alpha} n^\beta
n^\gamma-\left(M-\displaystyle{Q^2 \over
2r}\right)\displaystyle{2z \over r} n_{(l} \delta^{\alpha) 3}
\Biggr] -\displaystyle{a^3 \over
r^4}\Biggl[\Biggl\{6\left(M-\displaystyle{Q^2 \over
2r}\right)\displaystyle{z^2 \over r^2}\nonu
 \A \A +\displaystyle{Q^2
 \over 2r^3}\left(x^2+y^2\right)\Biggr\}n^\beta \epsilon_{\beta
3 (l}n^{\alpha)}-\left(M-\displaystyle{Q^2 \over
2r}\right)\displaystyle{2z \over r}n^\beta \epsilon_{\beta 3(l
}\delta^{\alpha) 3} \Biggr]+\displaystyle{a^4 \over 8r^5}
\Biggl[\Biggl\{\left(5M-\displaystyle{4Q^2 \over r}
\right)-\Biggl(14M\displaystyle{z^2 \over r^2}\nonu
\A \A +\left(63M-\displaystyle{40Q^2 \over r} \right)
\displaystyle{z^4 \over r^4} \Biggr) \Biggr\}n_l
n^\alpha-8\left\{\left(M-\displaystyle{Q^2 \over
r}\right)-\left(7M-\displaystyle{4Q^2 \over r} \right)
\displaystyle{z^2 \over r^2}\right\} \displaystyle{2z \over
r}n_{(l}\delta^{\alpha) 3}-4\Biggl\{\left(M-\displaystyle{Q^2
\over r}\right) \nonu
\A \A -\left(8M-\displaystyle{5Q^2 \over r} \right)
\displaystyle{z^2 \over r^2}+\left(7M-\displaystyle{4Q^2 \over r}
\right) \displaystyle{z^4 \over r^4}\Biggr\}{\delta_l}^\alpha
+4\left\{\left(M-\displaystyle{Q^2 \over
r}\right)-\left(9M-\displaystyle{5Q^2 \over r} \right)
\displaystyle{z^2 \over r^2}\right\}{\delta_{3l}}{\delta^{3
\alpha}} \Biggr]. \ea Also with this approximation the covariant
components of the metric tensor have the form
 \ba g_{0 0} \A=\A
1-\displaystyle{2 \over r}\left(M-\displaystyle{Q^2 \over
2r}\right)-\displaystyle{a^2 \over
r^3}\left[\left(M-\displaystyle{Q^2 \over
r}\right)-\left(3M-\displaystyle{2Q^2 \over
r}\right)\displaystyle{z^2 \over r^2}\right]\nonu
\A \A -\displaystyle{a^4 \over
4r^5}\left[\left(3M-\displaystyle{4Q^2 \over r}
\right)-\left(30M-\displaystyle{24Q^2 \over r}
\right)\displaystyle{z^2 \over r^2}+\left(35M-\displaystyle{24Q^2
\over r} \right)\displaystyle{z^4 \over r^4} \right],\nonu
g_{0 \alpha} \A=\A  -\left(M-\displaystyle{Q^2 \over
2r}\right){2n_\alpha \over r}+\displaystyle{2a \over
r^2}\left(M-\displaystyle{Q^2 \over 2r}\right)\epsilon_{\alpha
\beta 3} n^\beta+\displaystyle{2a^2 \over
r^3}\Biggl[\left\{2\left(M-\displaystyle{Q^2 \over
2r}\right)\displaystyle{z^2 \over r^2}+\displaystyle{Q^2 \over
4r^2}(x^2+y^2)\right\}n_\alpha\nonu
\A \A -\left(M-\displaystyle{Q^2 \over 2r} \right)\displaystyle{z
\over r}{\delta_\alpha}^3\Biggr]+\displaystyle{a^3 \over
r^4}\left[\left(M-\displaystyle{Q^2 \over
r}\right)-\left(5M-\displaystyle{3Q^2 \over
r}\right)\displaystyle{z^2 \over r^2}\right]\epsilon_{\alpha \beta
3} n^\beta+\displaystyle{a^4 \over
4r^5}\Biggl[3\Biggl\{\displaystyle{Q^2 \over
2r}+\left(8M-\displaystyle{7Q^2 \over r}\right) \displaystyle{z^2
\over r^2} \nonu
\A \A -\left(16M-\displaystyle{21Q^2 \over
2r}\right)\displaystyle{z^4 \over r^4}\Biggr\}n_\alpha
-2\left\{2\left(2M-\displaystyle{3Q^2 \over
2r}\right)-\left(12M-\displaystyle{7Q^2 \over
r}\right)\displaystyle{z^2 \over r^2}\right\}\displaystyle{z \over
r} {\delta_\alpha}^3 \Biggr],\nonu
g_{\alpha \beta} \A=\A -\delta_{\alpha
\beta}-2\left(\displaystyle{  n_\alpha n_\beta \over
r}-\displaystyle{2a \over r^2}\epsilon_{ \gamma 3 (\beta
}n_{\alpha)}n^\gamma \right)\left(M-\displaystyle{Q^2 \over
2r}\right)+\displaystyle{a^2 \over
r^3}\Biggl[\left\{M+\left(5M-\displaystyle{3Q^2 \over
r}\right)\displaystyle{z^2 \over r^2}\right\}n_\beta n_\alpha
\nonu
\A \A +2\left(M-\displaystyle{Q^2 \over 2r}\right)\epsilon_{
\epsilon \beta 3} {\epsilon_{\alpha \gamma}}^3n^\epsilon
n^\gamma-4\left(M-\displaystyle{Q^2 \over
2r}\right)\displaystyle{z \over r}n_{(\beta
}{\delta_{\alpha)}}^3\Biggr] -\displaystyle{a^3 \over
r^4}\Biggl[\Biggl\{12\left(M-\displaystyle{Q^2 \over
2r}\right)\displaystyle{z^2 \over r^2}\nonu
\A \A+\displaystyle{Q^2 \over
r^3}\left(x^2+y^2\right)\Biggr\}n^\gamma
n_{(\alpha}-4\left(M-\displaystyle{Q^2 \over 2r}\right)
\displaystyle{z \over r}n^\gamma
{\delta_{(\alpha}}^3\Biggr]\epsilon_{\beta)  \gamma 3
}+\displaystyle{a^4 \over 4r^5}
\Biggl[\Biggl\{\left(5M-\displaystyle{4Q^2 \over r}
\right)-\Biggl(14M\displaystyle{z^2 \over r^2}\nonu
\A \A +\left(63M-\displaystyle{40Q^2 \over r} \right)
\displaystyle{z^4 \over r^4} \Biggr) \Biggr\}n_\beta
n_\alpha-16\left\{\left(M-\displaystyle{Q^2 \over
r}\right)-\left(7M-\displaystyle{4Q^2 \over r} \right)
\displaystyle{z^2 \over r^2}\right\} \displaystyle{z \over
r}n_{(\beta}{\delta_{\alpha)}}^3-4\Biggl\{\left(M-\displaystyle{Q^2
\over r}\right)\nonu
\A \A -\left(8M-\displaystyle{5Q^2 \over r} \right)
\displaystyle{z^2 \over r^2}+\left(7M-\displaystyle{4Q^2 \over r}
\right) \displaystyle{z^4 \over r^4}\Biggr\}\delta_{\beta \alpha}
+4\left\{\left(M-\displaystyle{Q^2 \over
r}\right)-\left(9M-\displaystyle{5Q^2 \over r} \right)
\displaystyle{z^2 \over r^2}\right\}\delta_{\beta 3}\delta_{3
\alpha} \Biggr],
 \ea
 and the contravariant components of the metric tensor have the
 form
 \ba g^{0 0} \A=\A
1+\displaystyle{2 \over r}\left(M-\displaystyle{Q^2 \over
2r}\right)+\displaystyle{a^2 \over
r^3}\left[\left(M-\displaystyle{Q^2 \over
r}\right)-\left(3M-\displaystyle{2Q^2 \over
r}\right)\displaystyle{z^2 \over r^2}\right]\nonu
\A \A +\displaystyle{a^4 \over
4r^5}\left[\left(3M-\displaystyle{4Q^2 \over r}
\right)-\left(30M-\displaystyle{24Q^2 \over r}
\right)\displaystyle{z^2 \over r^2}+\left(35M-\displaystyle{24Q^2
\over r} \right)\displaystyle{z^4 \over r^4} \right],\nonu
g^{0 \alpha} \A=\A  g_{0 \alpha}, \nonu
g^{\alpha \beta} \A=\A -\delta^{\alpha
\beta}+2\left(\displaystyle{ n^\alpha n^\beta \over
r}-\displaystyle{2a \over r^2}\epsilon^{ \gamma 3 (\beta
}n^{\alpha)}n_\gamma \right)\left(M-\displaystyle{Q^2 \over
2r}\right)-\displaystyle{a^2 \over
r^3}\Biggl[\left\{M+\left(5M-\displaystyle{3Q^2 \over
r}\right)\displaystyle{z^2 \over r^2}\right\}n^\beta n^\alpha
\nonu
\A \A +2\left(M-\displaystyle{Q^2 \over 2r}\right)\epsilon^{
\epsilon \beta 3} {\epsilon^{\alpha \gamma}}_3n_\epsilon
n_\gamma-4\left(M-\displaystyle{Q^2 \over
2r}\right)\displaystyle{z \over r}n^{(\beta
}{\delta^{\alpha)}}_3\Biggr]+\displaystyle{a^3 \over
r^4}\Biggl[\Biggl\{12\left(M-\displaystyle{Q^2 \over
2r}\right)\displaystyle{z^2 \over r^2}\nonu
\A \A+\displaystyle{Q^2 \over
r^3}\left(x^2+y^2\right)\Biggr\}n_\gamma
n^{(\alpha}-4\left(M-\displaystyle{Q^2 \over 2r}\right)
\displaystyle{z \over r}n_\gamma
{\delta^{(\alpha}}_3\Biggr]\epsilon^{\beta)  \gamma 3
}-\displaystyle{a^4 \over 4r^5}
\Biggl[\Biggl\{\left(5M-\displaystyle{4Q^2 \over r}
\right)-\Biggl(14M\displaystyle{z^2 \over r^2}\nonu
\A \A +\left(63M-\displaystyle{40Q^2 \over r} \right)
\displaystyle{z^4 \over r^4} \Biggr) \Biggr\}n^\beta
n^\alpha-16\left\{\left(M-\displaystyle{Q^2 \over
r}\right)-\left(7M-\displaystyle{4Q^2 \over r} \right)
\displaystyle{z^2 \over r^2}\right\} \displaystyle{z \over
r}n^{(\beta}{\delta^{\alpha)}}_3-4\Biggl\{\left(M-\displaystyle{Q^2
\over r}\right)\nonu
\A \A -\left(8M-\displaystyle{5Q^2 \over r} \right)
\displaystyle{z^2 \over r^2}+\left(7M-\displaystyle{4Q^2 \over r}
\right) \displaystyle{z^4 \over r^4}\Biggr\}\delta^{\beta \alpha}
+4\left\{\left(M-\displaystyle{Q^2 \over
r}\right)-\left(9M-\displaystyle{5Q^2 \over r} \right)
\displaystyle{z^2 \over r^2}\right\}\delta^{\beta 3}\delta^{3
\alpha} \Biggr].
 \ea
The determinant of the metric tensor is \be g=-1.\ee Now evaluate
all the required components of the contorsion and the basic vector
using ((2) and (4)) neglecting terms beyond the fourth powers of
${\it a}$.

Since we are interested in evaluating the energy and momentum
density associated with the exterior solution (14), we evaluate
all the required components of the superpotential (24) neglecting
terms beyond the fourth order of $a$. Substituting (31), (32) and
all components of the contorsion and basic vector in (24) one gets
\ba {{\cal U}_0}^{0 \alpha} \A =\A {2n^\alpha \over \kappa r^2}
\Biggl[(M-\displaystyle{Q^2 \over 2r})+\displaystyle{a^2 \over
r^2}\left\{M(1-\displaystyle{5z^2 \over r^2})-\displaystyle{Q^2
\over r}(1-\displaystyle{3z^2 \over r^2})
\right\}+\displaystyle{a^4 \over 8r^4}\Biggl(M\Biggl\{9-126
\displaystyle{z^2 \over r^2}+ \displaystyle{189z^4 \over
r^4}\Biggr\} \nonu
\A \A -\displaystyle{Q^2 \over r}\Biggl\{ 12-\displaystyle{96z^2
\over r^2}+\displaystyle{120z^4 \over r^4}
\Biggr\}\Biggr)\Biggr]+{4z{\delta^\alpha}_3 \over \kappa r^3}
\Biggl[ \displaystyle{a^2 \over r^2}(M-\displaystyle{Q^2 \over
2r})+\displaystyle{a^4 \over 4r^4}\Biggl(M(9- \displaystyle{21z^2
\over r^2})\nonu
\A \A-6\displaystyle{Q^2 \over r}(1- \displaystyle{2z^2 \over
r^2})\Biggr)\Biggr]-{M n_\beta \epsilon^{\alpha \beta 3} \over
\kappa r^2}\Biggl(\displaystyle{a \over r}+\displaystyle{a^3 \over
r^3}(1-3\displaystyle{z^2 \over r^2})\Biggr)\; , \ea

 and
 \ba {{\cal U}_\alpha}^{0 \beta} \A =\A {1 \over 2
\kappa r^2}\Biggl[\displaystyle{Q^2 \over r}
{\delta_\alpha}^\beta+2\left(2M-\displaystyle{3Q^2 \over
2r}\right)n_\alpha n^\beta -\displaystyle{a \over
r}\Biggl\{\left(2M-\displaystyle{Q^2 \over
r}\right){\epsilon_{\alpha 3}}^{
\beta}+2\Biggl(3M-\displaystyle{2Q^2 \over r}\Biggr)
\epsilon_{\alpha \gamma 3} n^\gamma n^\beta \Biggr \} \nonu
\A \A+\displaystyle{a^2 \over
r^2}\Biggl\{2\left(M-\displaystyle{Q^2 \over
r}\right)\left({\delta_\alpha}^\beta-{\delta_\alpha}^3{\delta_3}^\beta\right)+
3\displaystyle{Q^2 \over
2r}{\delta_\alpha}^3{\delta_3}^\beta-\Biggl(24M-\displaystyle{35Q^2
\over 2r}\Biggr)\displaystyle{z^2 \over r^2} n_\alpha
n^\beta+\displaystyle{5Q^2 \over 2r}\epsilon_{\alpha \gamma
3}{\epsilon_\delta}^{3 \beta} n^\gamma n^\delta \nonu
\A \A +2\left(8M-\displaystyle{15Q^2 \over
2r}\right)\displaystyle{z \over r}
n^{(\beta}{\delta_{\alpha)}}^3\Biggr \} -\displaystyle{a^3 \over
r^3}\Biggl\{\left(M-\displaystyle{Q^2 \over
r}\right){\epsilon_{\alpha 3}}^\beta
+\Biggl[\left(5M-\displaystyle{6Q^2 \over r}\right)\nonu
\A \A -\left(35M-\displaystyle{24Q^2 \over r}\right)
\displaystyle{z^2 \over r^2} \Biggr]\epsilon_{\alpha \gamma
3}n^\gamma n^\beta+\left(10M-\displaystyle{6Q^2 \over
r}\right)\displaystyle{z \over r} n^\rho \; {\epsilon_{\alpha
\rho}}^3 \; {\delta^\beta}_3 -\left(5M-\displaystyle{3Q^2 \over
r}\right)\displaystyle{z^2 \over r^2} {\epsilon_{\alpha 3}}^\beta
\Biggr\} \nonu
\A \A+\displaystyle{a^4 \over
r^4}\Biggl\{\Biggl[\left(2M-\displaystyle{9Q^2 \over
4r}\right)-\left(54M-\displaystyle{357Q^2 \over
8r}\right)\displaystyle{z^2 \over
r^2}+\Biggl(168M-\displaystyle{126Q^2 \over
r}\Biggr)\displaystyle{z^4 \over
r^4}-\left(120M-\displaystyle{693Q^2 \over
8r}\right)\displaystyle{z^6 \over r^6}\Biggr]\nonu
\A
\A\Biggl({\delta_\alpha}^\beta-{\delta_\alpha}^3{\delta_3}^\beta\Biggr)+
15\displaystyle{Q^2 \over 8r}{\delta_\alpha}^3{\delta_3}^\beta
+\Biggl[\displaystyle{21Q^2 \over
8r}+\Biggl(48M-\displaystyle{189Q^2 \over
4r}\Biggr)\displaystyle{z^2 \over r^2}
-\Biggl(120M-\displaystyle{693Q^2 \over
8r}\Biggr)\displaystyle{z^4 \over r^4} \Biggr]\epsilon_{\alpha
\gamma 3}{\epsilon_\delta}^{3\beta}n^\gamma n^\delta
 \nonu
\A \A+2\Biggl[\Biggl(12M-\displaystyle{105Q^2 \over
8r}\Biggr)-\left(96M-\displaystyle{315Q^2 \over
8r}\right)\displaystyle{z^2 \over r^2}
 +\Biggl(120M-\displaystyle{693Q^2 \over
8r}\Biggr)\displaystyle{z^4 \over r^4} \Biggr] \displaystyle{z
\over r} n^{(\beta}{\delta_{\alpha)}}^3\nonu
\A \A +\Biggl[\Biggl(12M-\displaystyle{105Q^2 \over
8r}\Biggr)\displaystyle{z^2 \over
r^2}+\left(48M+\displaystyle{315Q^2 \over
8r}\right)\displaystyle{z^4 \over r^4}-\Biggl(120M
-\displaystyle{693Q^2 \over 8r}\Biggr)\displaystyle{z^6 \over
r^6}\Biggr] \delta_{\alpha 3} \delta^{\beta 3}\Biggr\}\Biggr].
 \ea

Now it is easy to calculate the components  of the energy-momentum
density. Those components we are interested in are given by \ba
{\tau_0}^0 \A=\A \displaystyle{Q^2 \over \kappa r^4} \left[1-
\displaystyle{a^2 \over  r^2}\left(2- \displaystyle{6 \over
r^2}(x^2+y^2)\right)+ \displaystyle{a^4 \over
r^4}\left(3-\displaystyle{24 \over r^2}(x^2+y^2)+\displaystyle{30
\over r^4}(x^2+y^2)^2\right) \right]\; ,\nonu
{\tau_\alpha}^0 \A=\A -\displaystyle{2 a Q^2 \epsilon_{\alpha
\beta 3} n^\beta \over \kappa r^5}\left(1+\displaystyle{3a^2 \over
r^2}(r^2-2z^2) \right). \ea

Further substituting (36) in (28) and then transforming it into
spherical polar coordinate one obtains \be E=\displaystyle{Q^2
\over \kappa r^6}\int_{r=R} \int \int \left(r^4+4a^2r^2-6a^2r^2
\cos^2 \theta+9a^4-36a^4 \cos^2 \theta+30a^4 \cos^4 \theta\right)
dr \sin \theta d\theta d\phi.\ee

Performing the above integration one gets the energy associated
with the exterior solution (14) in the form \be
{E(R)_{total}}^{exterior}=\displaystyle{Q^2 \over R}
\left(\displaystyle{1 \over 2}+\displaystyle{a^2 \over
3R^2}+\displaystyle{3a^4 \over 10R^4} \right).\ee

 Now we turn our attention to the angular momentum of solution
 (14). From (36) in (28) one can gets the components of the
 momentum density in the form
 \be P_\alpha= -\displaystyle{2 a Q^2 \epsilon_{\alpha \beta 3} n^\beta \over
\kappa r^5}\left(1+\displaystyle{3a^2 \over r^2}(r^2-2z^2)
\right)\; ,\ee from which one can shows that there is no spatial
momentum  associated with Kerr black hole. The components of the
angular momentum of a general-relativistic system is given by
\cite{VM} \be J_\alpha= \int \int \int (x_\beta P_\gamma-x_\gamma
P_\beta) d^3 x\; ,\ee where $\alpha$, $\beta$, $\gamma$ take
cyclic values 1,2,3. Using (39) in (40), transforming the
expressions into spherical polar coordinates
 \be J_\alpha=\int \int \int \displaystyle{2 a Q^2  \over
\kappa r^4} \left(r^2+3a^2(1-2\cos^2\theta \right)\sin^3\theta dr
d\theta d\phi\; ,\ee performing the above integration one gets \be
J_\alpha=-2aQ^2 \epsilon_{12\alpha} \left[\displaystyle{1 \over
3r}+\displaystyle{a^2 \over 5r^3} \right]_{R_1}^{R_2}.\ee
\newsection{Main results and discussion}

In this paper we have studied the coupled equations of the
gravitational and electromagnetic fields  in the tetrad theory of
gravitation, applying the most general  tetrad (13) with sixteen
unknown function of $\rho$ and $\phi$ to the field equations
(7)$\sim$(9). Exact analytic solution is obtained (14). This
solution (14) is a general solution from which one can generates
the other solutions (Schwarzschild, Reissner-Nordstr$\ddot{o}$m
and  Kerr spacetimes) by an appropriate choice of the arbitrary
functions of the tetrad (13).

It was shown by M\o ller \cite{Mo26} that the tetrad description
of the gravitational field allows a more satisfactory treatment of
the energy-momentum complex than does general relativity.
Therefore, we have  used the superpotential (24) to calculate the
energy and spatial momentum (28). Because the definition of energy
(28) requires its evaluation in Cartesian coordinate, the
calculations without any approximation is obviously very lengthy.
Moreover the intrinsic rotation parameter $a$ is quantitatively
very small for most physical situations and so for our convenience
we keep terms containing powers of $a$ up to the fourth order,
(i.e., $a^4$). With this approximation we have calculated the
components of a covariant and contravariant tetrad fields (29) and
(30), a covariant and contravariant components of metric tensor
(31) and (32). Calculating all the necessary components of the
contorsion and basic vector ((2) and (4))  and using (31) and (32)
in (28) we have obtained the expression of the exterior energy of
solution (14) till the fourth order. When rotation as well as
charge parameters both are considered we get an additional terms
$\displaystyle{Q^2 \over R} \left(\displaystyle{1 \over
2}+\displaystyle{a^2 \over 3R^2}+\displaystyle{3a^4 \over 10R^4}
\right)$ which is the energy of the exterior magnetic field due to
the rotation of the charged object. The asymptotic value of the
total gravitational mass is the Schwarzschild mass and therefore,
the energy associated with the axially symmetric solution (14)
contained in a sphere of radius $R$ is \be
E(R)=M-\displaystyle{Q^2 \over R} \left(\displaystyle{1 \over
2}+\displaystyle{a^2 \over 3R^2}+\displaystyle{3a^4 \over 10R^4}
\right). \ee  Switching off the rotation parameter, $(i.e., a=0 )$
the energy associated will be the same as that of
Reissner-Nordstr$\ddot{o}$m metric \cite{Vc,Tp}. Setting  the
charge parameter $Q$ to be  equal zero, i.e., in the case of a
Kerr black hole, it is clear from the expression (43) that there
is no energy contained by the exterior of the Kerr black hole and
hence the entire energy is confined to its interior only.

Using (36) we have calculated the spatial momentum neglecting
terms beyond the fourth order. Substituting (39) in (40) we have
calculated the angular momentum distribution due to the
electromagnetic field  which as is clear from (42) that it depends
mainly on the rotation of the charged object. We got only the Z
component of the angular momentum which is consistence with the
structure of the charged axially symmetric solution (14) that
describe the exterior field of a charged object rotating about the
Z axis. As is clear from (42) that the angular momentum depends on
the even powers of the charge parameter and odd powers of the
rotation parameter which means that the direction of the angular
momentum vector depends on the direction of the rotation
parameter.

Aguirregabiria et al. have computed the distributions for the
Kerr-Newman and Bonnor-Vaidya metrics and found a reasonable
results. Their calculations are performed  without any
approximations in Kerr-Schild Cartesian coordinates \cite{ACV}.
Here we have calculated the distribution of the charged axially
symmetric solution in the Cartesian coordinate. The calculations
without any approximation will be very lengthy therefore, the
formula of energy and angular-momentum have been given up to
$O(a^4)$. The calculated energy (38) is in a good agreement with
Aguirregabiria at el. up to $O(a^4)$.

In a summary, the tetrad theory of gravitation creates only {\rm a
unique  definition  of the energy-momentum complex} \cite{Mo8}
which we have used.  The advantage of this definition that it
gives a satisfactory results for the calculations of energy and
spatial momentum related to the charged axially symmetric solution
(14). In contrary to general relativity, in which it possesses
many definitions of the energy-momentum complexes. The results of
energy and spatial momentum of  some of these definitions agree
and the others do not for the same spacetime \cite{Vs,Vs1}. This
is clear from what has been done by Virbhadra, who calculated the
energy of Kerr-Newman metric up to O$(a^3)$, using many
definitions of the energy-momentum complexes "which depend mainly
on the metric". Some of these definitions are in agreement  in
their results but others do not \cite{Vs,Vs1}. Here the
calculations have done up to $(a^4)$ using definition (24) which
depend mainly on the tetrad field ${b^i}_\mu$.

\bigskip
\bigskip
\centerline{\Large{\bf Acknowledgements}}

The author would like to thank Professor I.F.I. Mikhail; Ain Shams
University Egypt, for his  stimulating discussions, Professor T.
Shirafuji Saitama university, Japan and Professor O. Shalabiea
Cairo University, Egypt.

\end{document}